\documentclass[prl,twocolumn,aps]{revtex4-1}
\usepackage{graphicx}
\usepackage{comment}
\usepackage{bm}
\usepackage{amsmath,amssymb} 
\usepackage{epstopdf}
\usepackage{verbatim}
\usepackage{float}

\begin{document}

\title{Do the repulsive and attractive pair forces play separate roles for the physics of liquids?}
\author{Lasse B{\o}hling, Arno A. Veldhorst, Trond S. Ingebrigtsen, Nicholas P. Bailey, Jesper S. Hansen, S{\o}ren Toxvaerd, Thomas B. Schr{\o}der, and Jeppe C. Dyre}
\email{dyre@ruc.dk}
\affiliation{DNRF Centre ``Glass and Time'', IMFUFA, Department of Sciences, Roskilde University, Postbox 260, DK-4000 Roskilde, Denmark}
\date{\today}

\begin{abstract}
According to standard liquid-state theory repulsive and attractive pair forces play distinct roles for the physics of liquids. This paradigm is put into perspective here by demonstrating a continuous series of pair potentials that have virtually the same structure and dynamics, although only some of them have attractive forces of significance. Our findings reflect the fact that the motion of a given particle is determined by the total force on it, whereas the quantity usually discussed in liquid-state theory is the individual pair force.
\end{abstract}

\maketitle

A liquid is held together by attractions between its molecules. On the other hand it is very difficult to compress a liquid because the molecules strongly resist closely approaching each other. These facts have been well known for a long time, and today it is conventional wisdom that the repulsive and the attractive forces play distinct roles for the physics of liquids. The repulsive forces, which ultimately derive from the Fermi statistics of electrons, are harsh and short ranged. According to standard theory these forces are responsible for the structure and, in particular, for reducing considerably the liquid's entropy compared to that of an ideal gas at the same density and temperature. The attractive forces, on the other hand, are long ranged and weaker. These forces, which derive from induced dipolar interactions, reduce the pressure and energy compared to that of an ideal gas at the same density and temperature. We argue below that this physical picture, though quite appealing, overemphasizes the individual pair forces and does not provide a full understanding because it does not relate directly to the total force on a given particle.

The traditional understanding of the liquid state is based on pioneering works by Frenkel, Longuet-Higgens and Widom, Barker and Henderson, and Weeks, Chandler, and Andersen (WCA), and many others \cite{books,papers}. The basic idea is that the attractions may be regarded as a perturbation of a Hamiltonian based on the repulsive forces, the physics of which is usually well represented by a hard-sphere reference system \cite{zwa54}. Perturbation theories based on this picture \cite{books,papers,zwa54,zho09} are standard for calculating simple liquids' thermodynamics and structure as quantified, e.g., by the radial pair distribution function $g(r)$. We do not question the usefulness of perturbation theories, but will argue from theory and simulations that the repulsive and the attractive pair forces do not always play clearly distinguishable roles for the structure and dynamics of simple liquids.

\begin{figure}[H]
  \centering
  \includegraphics[width=80mm]{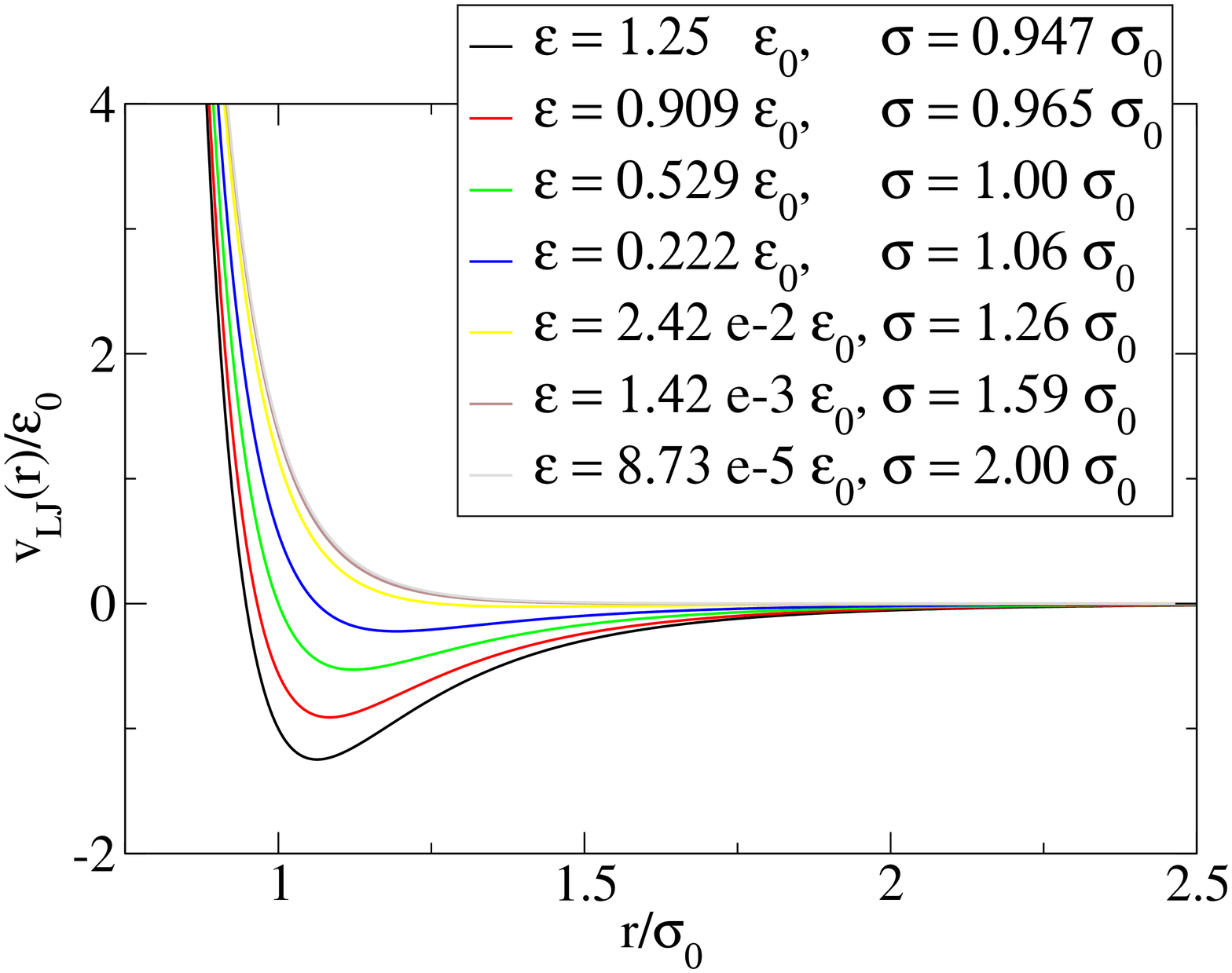}
  \caption{Lennard-Jones pair potentials $v_{\rm LJ}(r)=4\varepsilon[(r/\sigma)^{-12}-(r/\sigma)^{-6}]$ predicted to give the same physics at the state point $(\rho,T)=(1,1)$ using the units where $\epsilon_0=\sigma_0=1$ and $k_B=1$. Visually, these potentials have little in common; in particular, they have very different contributions from attractive forces. The pair potentials were constructed analytically using the isomorph theory, as detailed in the text after Fig. \ref{fig2}.}
  \label{fig1}
\end{figure}

This point is illustrated in the simplest possible way by studying systems of Lennard-Jones (LJ) particles. The LJ pair potential is given by $v_{\rm LJ}(r)=4\varepsilon[(r/\sigma)^{-12}-(r/\sigma)^{-6}]$. It is plotted in Fig.~\ref{fig1} for a number of different choices of the parameters $\epsilon$ and $\sigma$. In the following we adopt the unit system where $\epsilon_0=\sigma_0=1$ and $k_B=1$. We use the same unit system for all the potentials. Consider a simulation of the potential with $(\epsilon,\sigma)=(1.25,0.947)$ at the state point $(\rho\equiv N/V,T)=(1,1)$. Clearly, it would lead to exactly the same structure and dynamics (after appropriate rescaling) doing a simulation of the potential with $(\epsilon,\sigma)=(8.73\cdot10^{-5},2.0)$ at the temperature, $T=8.73\cdot10^{-5}/1.25$, and density, $\rho=(0.947/2.00)^3$ -- this simply reflects the fact that the physics is determined by two dimensionless parameters $T/\epsilon$ and $\sigma^3\rho$. We show below, however, that in addition to this trivial fact, the two potentials also gives (to a good approximation) the same structure and dynamics when \emph {both} potentials are investigated at the state point $(\rho,T)=(1,1)$. In fact, all the potentials in Fig. \ref{fig1} were chosen so that they give virtually the same structure and dynamics at the state point $(\rho,T)=(1,1)$. The paper mainly focuses on this state point, but at the end of the paper results for a few other state points are also given, confirming the findings at $(\rho,T)=(1,1)$. 

The potentials of Fig. \ref{fig1} all have attractive forces, but for some of the potentials the attractive forces are entirely insignificant. To show that these potentials nevertheless have the same structure and dynamics, $NVT$ computer simulations of systems of 1,000 particles were performed using the RUMD software that runs on graphical processing units \cite{rumd}.

\begin{figure}[H]
  \centering
  \includegraphics[width=80mm]{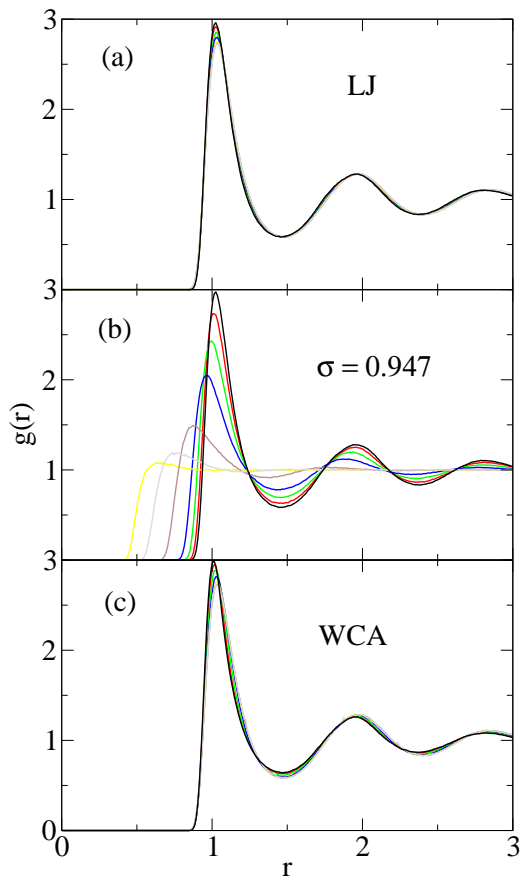}
  \caption{Radial distribution functions at the state point $(\rho,T)=(1,1)$ for different sets of potentials. (a) The LJ pair potentials of Fig. \ref{fig1}.
(b) A series of LJ pair potentials with fixed $\sigma$ parameter and the $\epsilon$-values listed in Fig. \ref{fig1}.
(c) Results for the series of Weeks-Chandler-Andersen (WCA) potentials corresponding to the LJ potentials of Fig. \ref{fig1}.}
  \label{fig2}
\end{figure}

Figure \ref{fig2}(a) shows the radial distribution function $g(r)$ for the seven LJ pair potentials of Fig. \ref{fig1} at the state point $(\rho,T)=(1,1)$. For comparison, simulations at the same state point are shown in Fig.  \ref{fig2}(b) for seven potentials with the same $\varepsilon$ variation, but fixed $\sigma=0.947$. Figure \ref{fig2}(c) shows the radial distribution functions at the state point $(1,1)$ for the pair potentials of Fig. \ref{fig1} cut off according to the Weeks-Chandler-Andersen (WCA) recipe, i.e., by cutting the potentials at their minima and shifting them to zero there.

\begin{figure}[H]
  \centering
  \includegraphics[width=80mm]{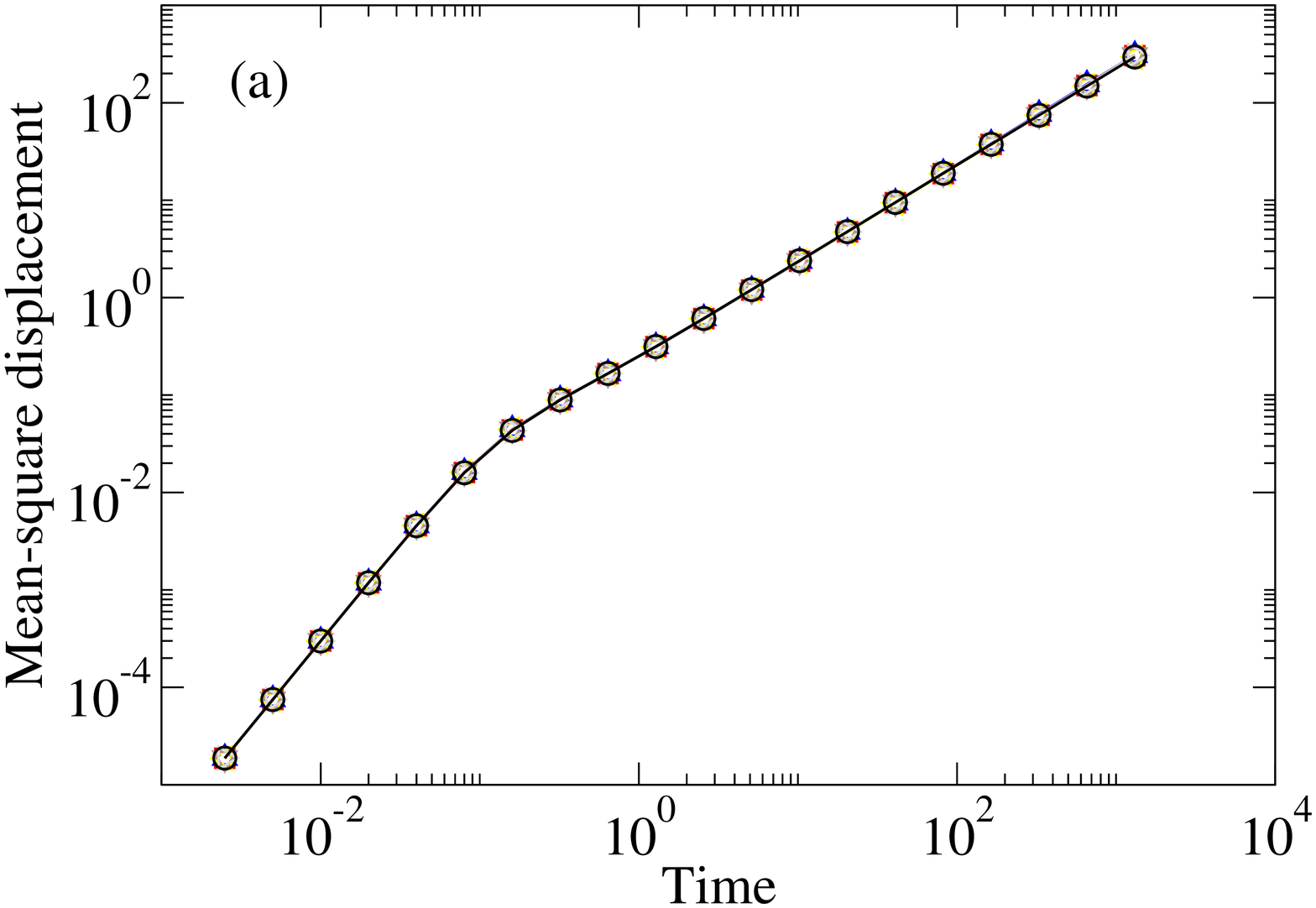}
  \includegraphics[width=80mm]{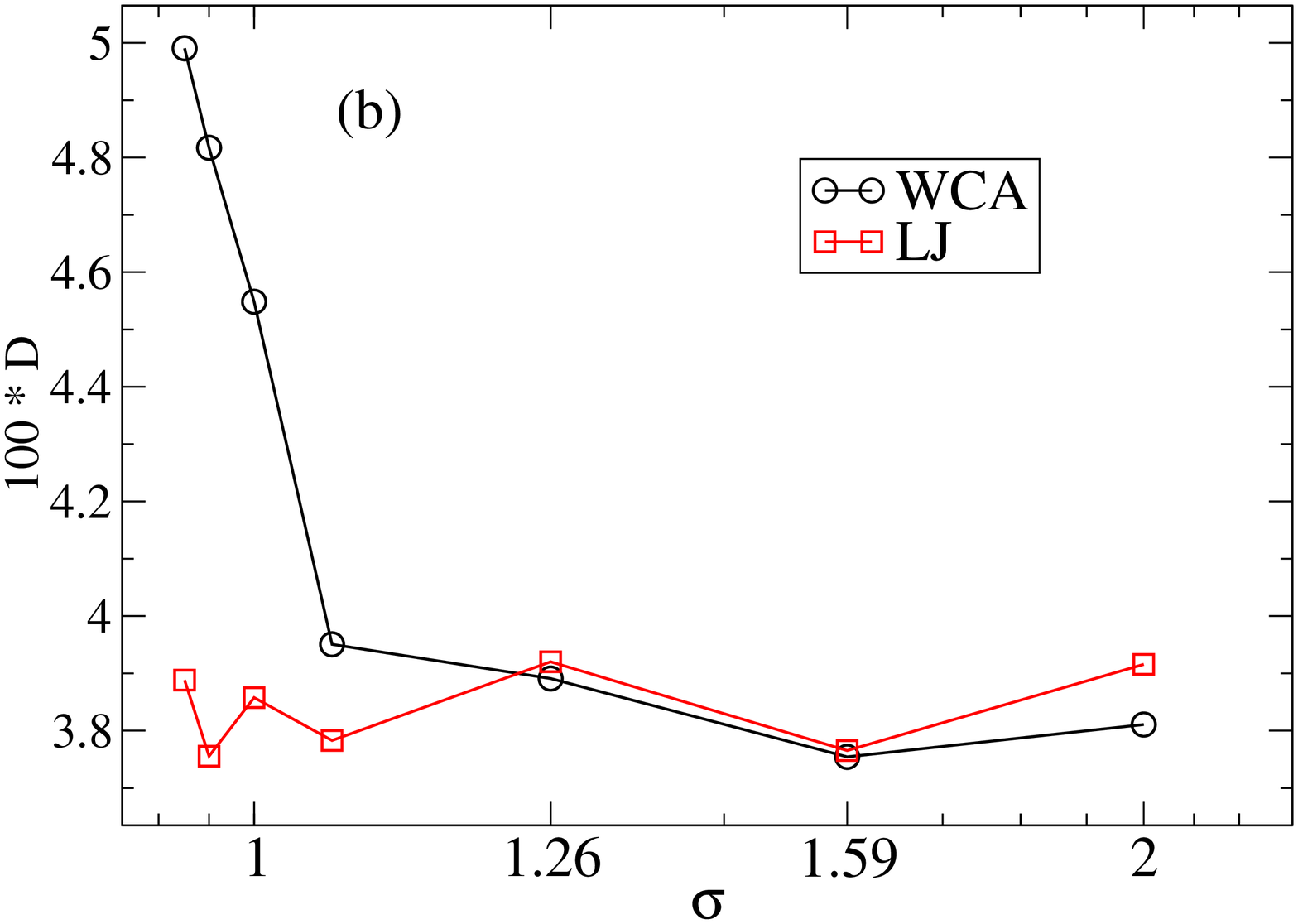}
  \caption{(a) The mean-square displacement for the LJ pair potentials of Fig. \ref{fig1} at the state point $(\rho,T)=(1,1)$.
(b) Reduced diffusion constants as functions of $\sigma$ for the full potentials of Fig. \ref{fig1} (red) and for the WCA versions of the potentials (black). At high $\sigma$ the WCA results are accurate because these potentials are almost purely repulsive.}
  \label{fig2ny}
\end{figure}

Figure \ref{fig2ny} shows results for the dynamics, with (a) giving the mean-square displacement for the seven potentials of Fig. \ref{fig1}. Figure \ref{fig2ny}(b) compares the results for the diffusion constants with those of WCA simulations. 

By the Henderson uniqueness theorem \cite{hen74} the pair potentials of Fig. \ref{fig1} cannot have exactly the same pair distribution functions. Based on Figs. 2 and 3 we see that, nevertheless, the´ LJ systems lead to very similar structure as well as  very similar dynamics. In fact, both structure and dynamics among the potentials of Fig. \ref{fig1} are closer to each other than to the WCA versions of the potentials. If the WCA approximation gave a faithful representation of the physics involved, this would not be the case. 

How were the pair potentials of Fig. \ref{fig1} determined and why do they have the almost same structure and dynamics? The starting point is the existence of isomorphs in the phase diagram of liquids with strong correlations between $NVT$ virial and potential-energy equilibrium fluctuations \cite{IV,scl} (which we recently argued provides a useful definition of a simple liquid \cite{prx}). Two state points with density and temperature $(\rho_1,T_1)$ and $(\rho_2,T_2)$ are termed isomorphic \cite{IV} if all pairs of  physically relevant microconfigurations of the two state points, which trivially scale into one another, i.e., $\rho_1^{1/3}{\bf r}_i^{(1)}=\rho_2^{1/3}{\bf r}_i^{(2)}$ for all particles $i$, have proportional configurational Boltzmann factors: $\exp[-U({\bf r}_1^{(1)},...,{\bf r}_N^{(1)}) /k_BT_1]=C_{12}\exp[-U({\bf r}_1^{(2)},...,{\bf r}_N^{(2)})/k_BT_2]$. LJ systems are strongly correlating and thus have isomorphs to a good approximation \cite{scl}. The invariance of the canonical probabilities of scaled configurations along an isomorph has several implications \cite{IV}: Excess entropy and isochoric specific heat are both isomorph invariant, the dynamics in reduced units are invariant for both Newtonian and Brownian equations of motion, reduced-unit static density correlation functions are invariant, a jump between two isomorphic state points takes the system instantaneously to equilibrium, etc. For Newtonian dynamics using reduced units corresponds to measuring length in units of $\rho^{-1/3}$, time in units of $\rho^{-1/3}\sqrt{m/k_BT}$ where $m$ is the particle mass, and energy in units of $k_BT$. Thus the reduced particle coordinates are defined by ${\bf \tilde r}_i=\rho^{1/3}\bf r_i$.

An isomorph was generated using the recently derived result \cite{ing12} that liquids with good isomorphs have simple thermodynamics in the sense that the temperature is a product of a function of excess entropy per particle $s$ and a function of density, 

\begin{equation}\label{eos}
T\,=\,f(s)h(\rho)\,.
\end{equation}
The function $h(\rho)$ inherits the analytical structure of the pair potential in the sense that if the latter is given by the expression $v(r)=\sum_n v_nr^{-n}$, then $h(\rho)=\sum_n C_n \rho^{n/3}$ in which each term corresponds to a term in the pair potential \cite{ing12}. Since $h(\rho)$ is only defined within an overall multiplicative constant, one can write for the LJ pair potential 

\begin{equation}\label{h}
h(\rho)\,=\,
\alpha \rho^4 +(1-\alpha)\rho^2\,.
\end{equation}
The constant $\alpha$ was determined from simulations at the reference state point $(\rho,T)=(1,1)$ for $\varepsilon = 1.25$ and $\sigma=0.947$, which is a typical liquid state point of the LJ system. This was done by  proceeding as follows \cite{beyond}. We have previously \cite{IV,ing12} derived the identities

\begin{equation}\label{gamma}
\gamma \equiv  \left(\frac{\partial\ln T}{\partial\ln\rho}\right)_{S_{\rm ex}}
=\frac{d\ln h}{d\ln\rho}
=\frac{\langle\Delta W\Delta U \rangle}{\langle(\Delta U)^2\rangle}\,,
\end{equation}
in which $W$ is the virial, $U$ the potential energy, and the angular brackets denote $NVT$ ensemble averages. Combining Eqs. (\ref{h}) and (\ref{gamma}) with the simulation results for the fluctuations of $W$ and $U$ leads to $\alpha=\gamma/2-1=1.85$.

An isomorph is a set of state points with almost invariant structure and dynamics in reduced units \cite{IV}. Via appropriate rescaling, however, an isomorph can be interpreted differently: as defining a set of {\it different} LJ pair potentials that give invariant properties at the {\it same state point}. These are simply two different ways of looking at an invariant Boltzmann factor: Equation (\ref{eos}) implies that $\exp \left( - {U(\rho^{-1/3}\tilde{\bf {r}}_1, \ldots, \rho^{-1/3}\tilde {\bf r}_N)} /{[f(s)h(\rho)]} \right) =\exp\left( - [1/f(s)] \sum_{i<j} {v_{\rm LJ}\left(\rho^{-1/3} \tilde r_{ij}\right)}/{h(\rho)} \right) $ where $r_{ij}$ is the distance between particles $i$ and $j$. Along an isomorph $f(s)$ is a constant; if we consider the isomorph which includes the state point $\rho=T=1$, then given the normalization of Eq. (\ref{h}) we have $f(s)=1$. The shift in interpretation now comes by noticing that the same Boltzmann factor is obtained by considering a configuration at unity density and unity temperature and a family of {\it isomorphic pair potentials} $v^{d}_{\rm LJ} (r) \equiv v_{\rm LJ}\left(d^{-1/3} r\right) / h(d)$, where we have dropped the tilde from positions and replaced $\rho$ with $d$ to emphasize the shift in perspective. These pair potentials are still LJ potentials, but with different energy and length parameters; these are the potentials plotted in Fig. \ref{fig1}.

\begin{figure}[H]
  \centering
  \includegraphics[width=80mm]{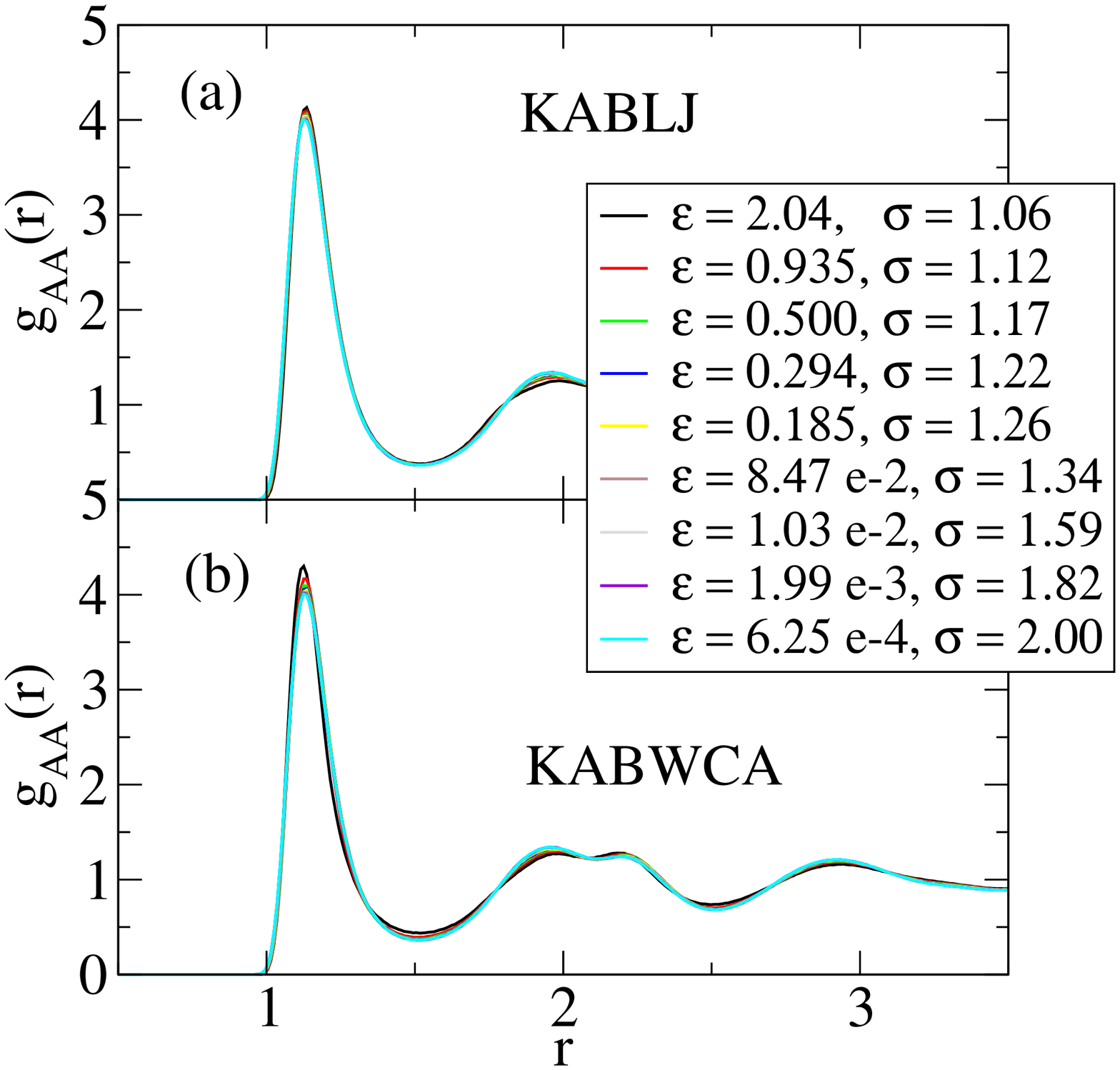}
  \includegraphics[width=80mm]{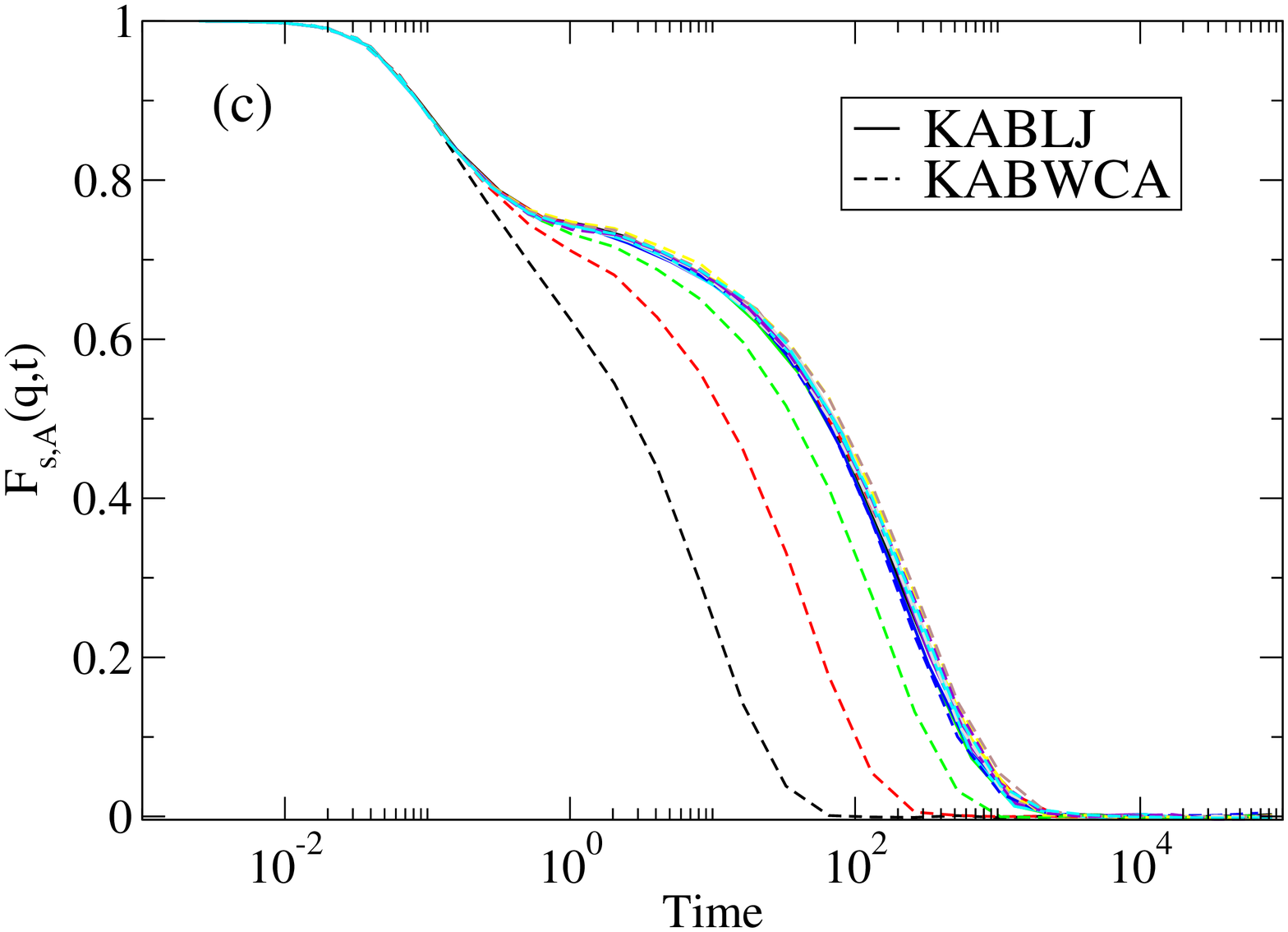}
  \includegraphics[width=80mm]{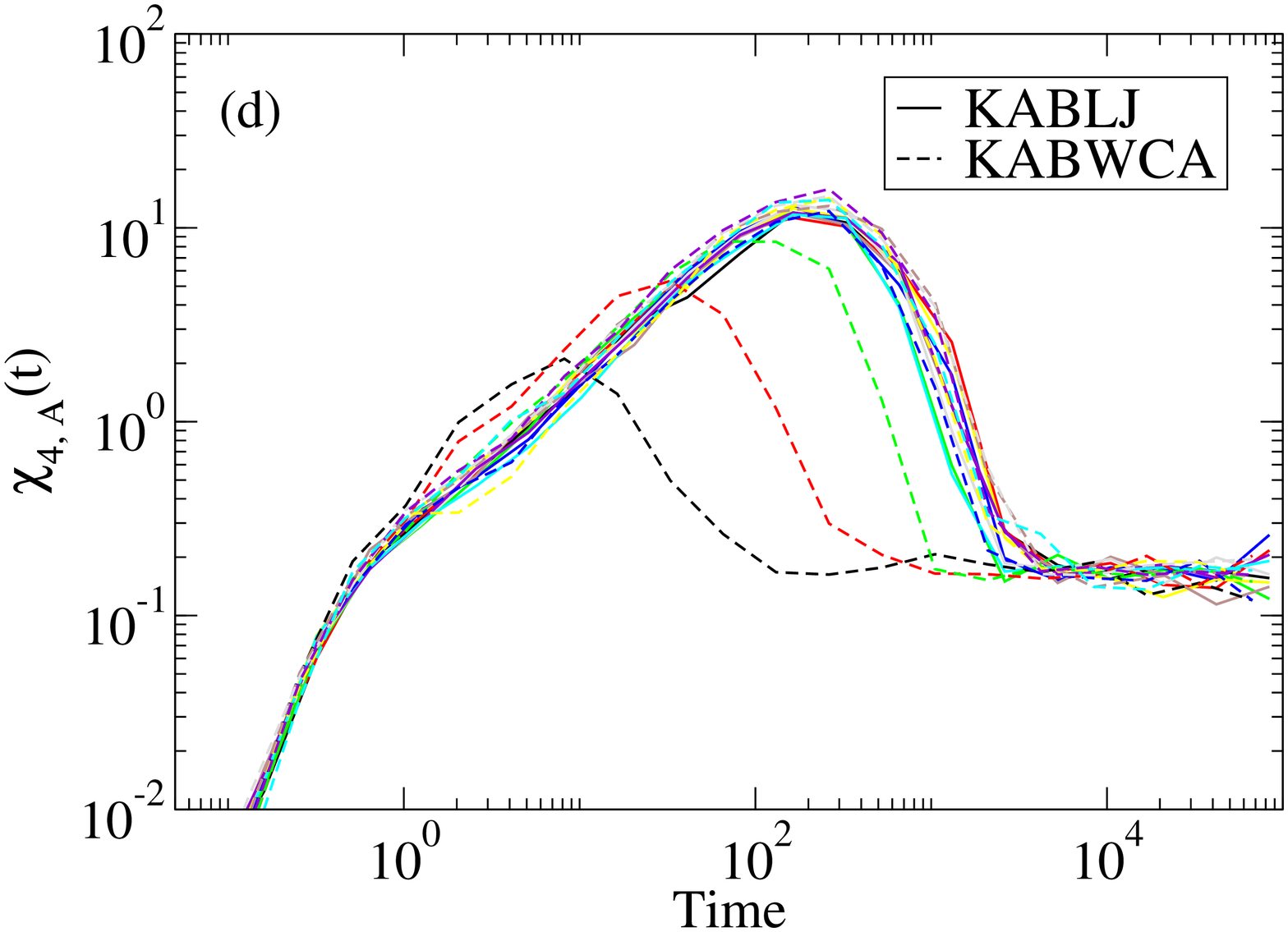}
  \caption{(a) The AA particle radial distribution function of the Kob-Andersen binary Lennard-Jones (KABLJ) mixture for a family of isomorphic pair potentials similar to those of Fig. \ref{fig1}.
(b) The AA particle radial distribution function of the KABLJ mixture with the corresponding WCA potentials.
(c) The A particle incoherent intermediate scattering function for the same family of potentials as a function of time at the wavevector defined from the maximum of $g(r)$ (full curves). The full dotted lines show the WCA predictions \cite{ber}.
(d) The function $\chi_4(t)$ for the A particles for the same pair potentials (full curves) and the WCA predictions (dashed lines).}
  \label{fig3}
\end{figure}

The single-component LJ system does not have a broad dynamic range because it cannot be deeply supercooled. To test the robustness of the predicted invariance of the physics for families of ``isomorphic'' pair potentials, we simulated also the Kob-Andersen binary LJ (KABLJ) mixture \cite{kablj}, which is easily supercooled into a highly viscous state. For this system the constant $\alpha=1.29$ was identified from simulations of 1,000 particles at the reference state point $(\rho,T)=(1.60,2.00)$, using again Eq. (\ref{gamma}). From the function $h(\rho)$ a family of isomorphic equivalent pair potentials was generated that looks much like those of Fig. \ref{fig1}; in particular, some of them have a vanishingly small attraction.

Figure \ref{fig3}(a) shows the AA particle radial distribution functions for these different pair potentials and Fig. \ref{fig3}(b) shows the same quantity for the WCA version of the potentials. Figure \ref{fig3}(c) shows the A particle incoherent intermediate scattering function and, with dashed lines, simulations of the corresponding WCA systems. Even though the WCA approximation has the correct repulsive forces, its physics differs considerably from the isomorphic pair potentials as noted already by Berthier and Tarjus \cite{ber}. We also calculated $\chi_4(t)$, a measure of dynamic heterogeneities. The results shown in Fig. \ref{fig3}(d) are more noisy, but confirm the predicted invariance of the dynamics for the different pair potentials. The corresponding WCA results are shown with dashed lines.

\begin{figure}[H]
  \centering
  \includegraphics[width=80mm]{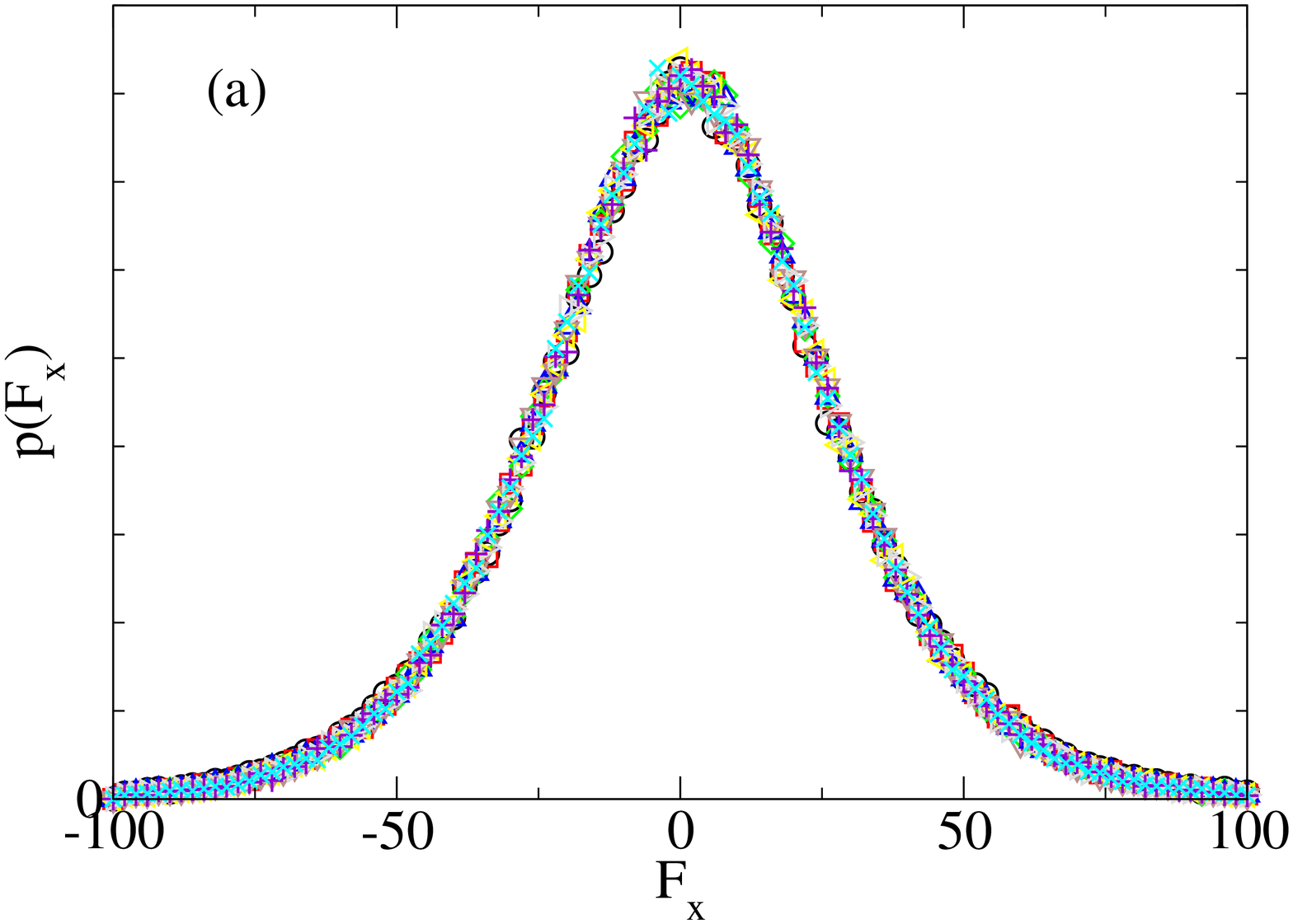}
  \includegraphics[width=80mm]{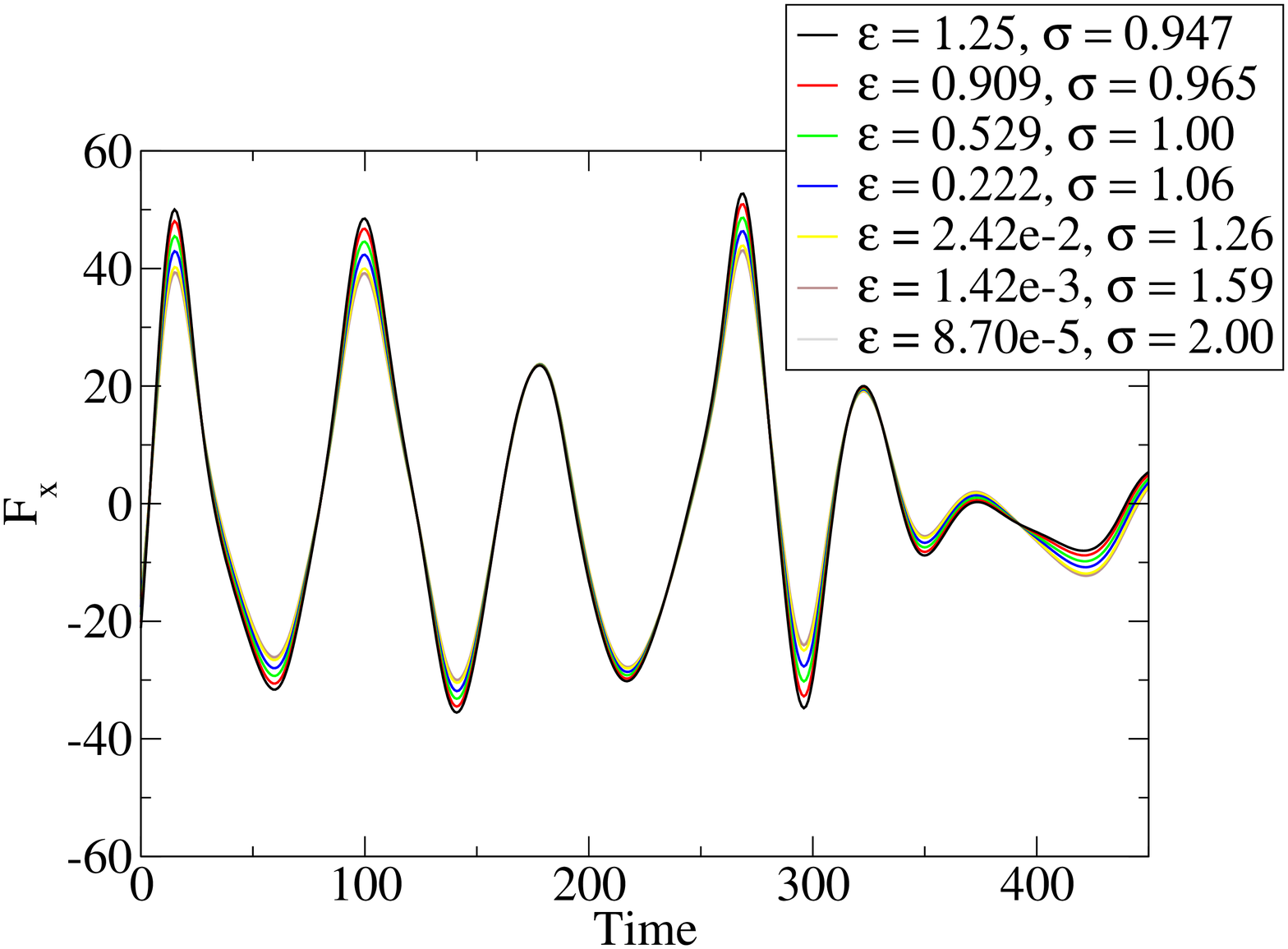}
  \caption{(a) Probability distribution of x-components of the total forces on individual particles, $p(F_x)$, for the different single-component LJ potentials of Fig. \ref{fig1} at the state point $(\rho,T)=(1,1)$.
(b) Snapshot of the $x$-component of the force $F_x$ on one particle as a function of time. The system simulated is defined by $\varepsilon=1.25$ and $\sigma=0.947$, and $F_x$ was subsequently evaluated for the same series of configurations for the six other potentials. These figures show that, even though the pair potentials are quite different, the forces are virtually identical except at the extrema.}
  \label{fig4}
\end{figure}

It would require extraordinary abilities to know from inspection of Fig. \ref{fig1} that these pair potentials have virtually the same structure and dynamics. The potentials have neither the repulsive nor the attractive terms in common, so why is it that they have such similar behavior? The answer is that they result in virtually the same forces (Fig. \ref{fig4}). The force on a given particle is the {\it sum} of contributions from (primarily) its nearest neighbors, so plotting merely the pair potential can be misleading. We conclude that by reference to the pair potential alone, one cannot identify separate roles for the repulsive and the attractive forces in a many-particle system. There simply are no ``repulsive'' and ``attractive'' forces as such.

\begin{figure}[H]
  \centering
  \includegraphics[width=80mm]{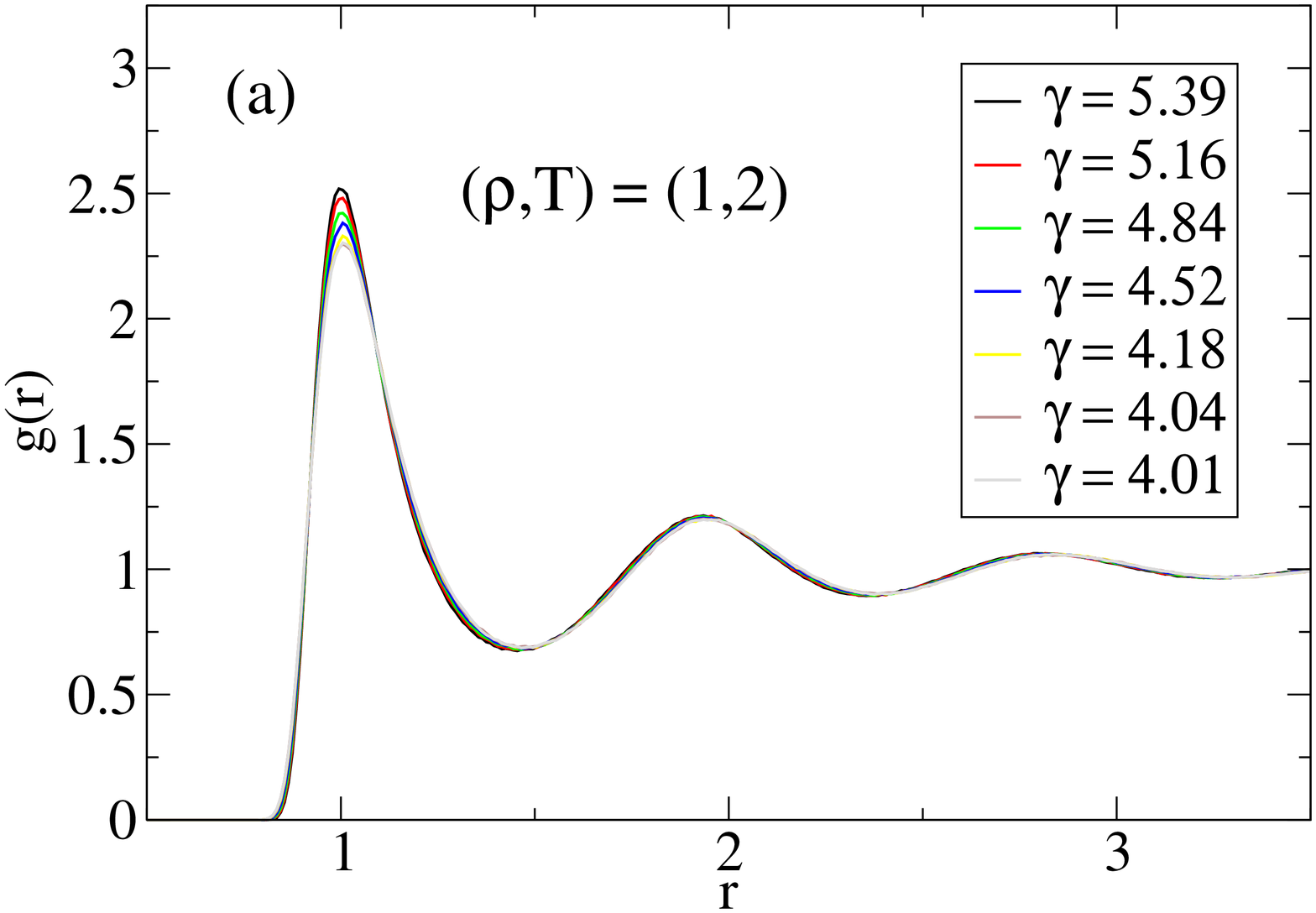}
  \includegraphics[width=80mm]{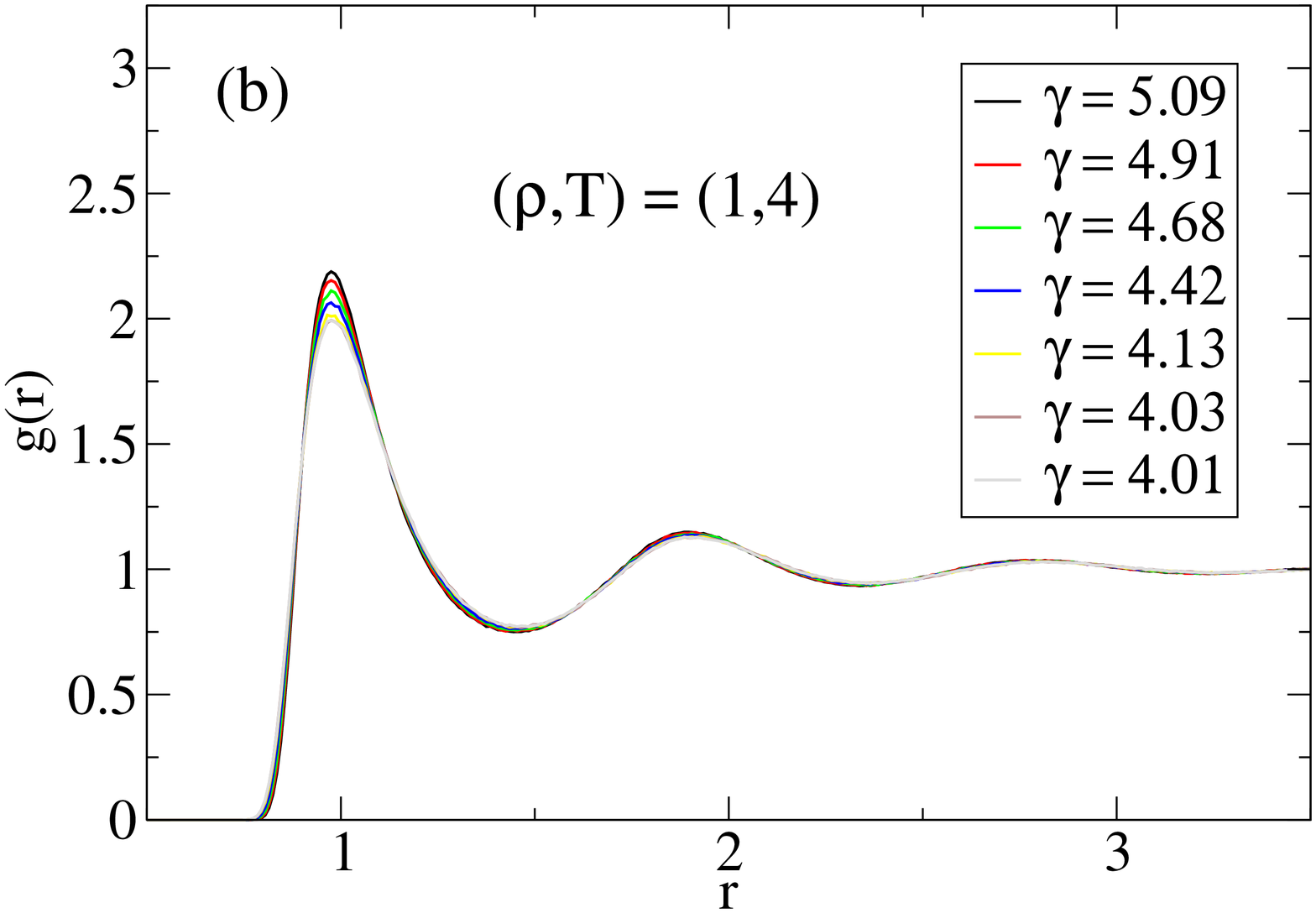}
  \includegraphics[width=80mm]{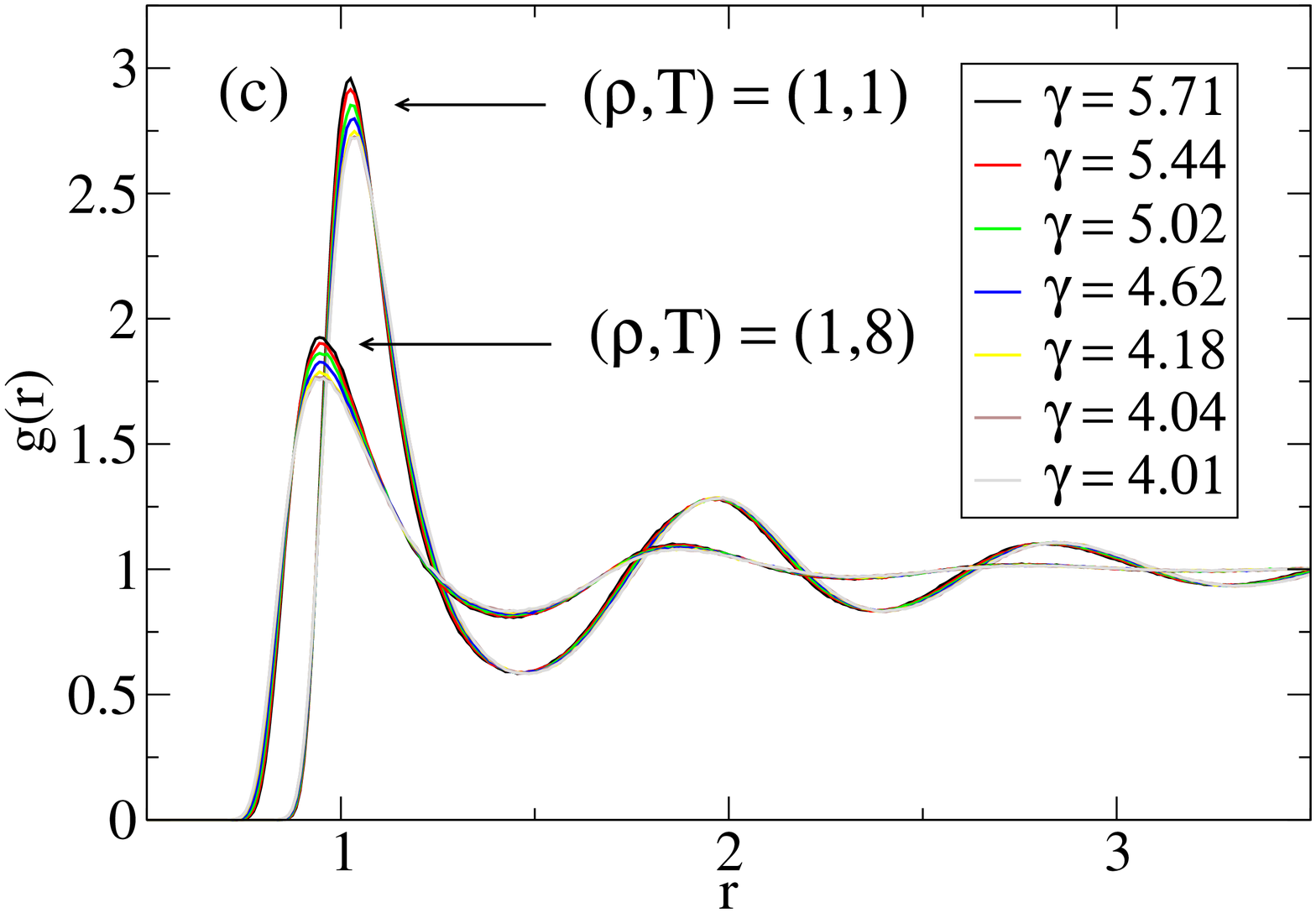}
  \caption{Radial distribution functions for the potentials of Fig. \ref{fig1} at other state points than the state point $(\rho,T)=(1,1)$ studied above. For reference we give in each subfigure the value of $\gamma$ defined in Eq. (\ref{gamma}).
(a) $(\rho,T)=(1,2)$; 
(b) $(\rho,T)=(1,4)$; 
(c) $(\rho,T)=(1,8)$ -- the $\gamma$'s reported in this subfigure are those of the state point $(\rho,T)=(1,1)$.}
  \label{fig5}
\end{figure}

The above reported simulations focussed for each system on one particular state point. If the potentials in Fig. \ref{fig1} are to be regarded as equivalent with respect to structure and dynamics, however, one should test also other state points. We have done this briefly for the single-component LJ liquid of Fig. \ref{fig1}. The results are shown in Fig. \ref{fig5}. Clearly, the degree of similarity observed at the state point $(\rho,T)=(1,1)$ is maintained also for the other state points (for comparison Fig. \ref{fig5}(c) reproduces the $(\rho,T)=(1,1)$ results from Fig. \ref{fig2}(a)).

What are the implications of the above results? For liquid state perturbation theory the WCA theory is rightfully renowned for its ability to make semi-analytic predictions for thermodynamic properties of simple liquids. The focus of liquid state theory has moved on, however, in part because modern computers make it straightforward to simulate the kinds of liquids for which WCA theory can make accurate predictions. We do not claim to have a better way to do perturbation theory in the sense of WCA. While WCA theory is based upon an {\em assumed} equivalence between two potentials differing by the removal of attractions, the present work describes an {\em observed} equivalence between apparently quite different potentials. This observation will not facilitate perturbation theory, but it could potentially be useful as a check on perturbation theories and other theories of the liquid state, for example density functional theory; such theories should be consistent with the observed invariance as the parameters of the potential are changed.

\acknowledgments 
The centre for viscous liquid dynamics ``Glass and Time'' is sponsored by the Danish National Research Foundation (DNRF).

\end{document}